# Helping students become proficient problem solvers Part I: A brief review

Alexandru Maries[1] and Chandralekha Singh[2]
[1]Department of Physics, University of Cincinnati, Cincinnati, OH 45221, USA
[2]Department of Physics and Astronomy, University of Pittsburgh, Pittsburgh, PA, 15260, USA

Understanding issues involved in expertise in physics problem solving is important for helping students become good problem solvers. In part 1 of this article, we summarize the research on problem-solving relevant for physics education across three broad categories: knowledge organization, information processing and cognitive load, and metacognition and problem-solving heuristics. We also discuss specific strategies discussed in the literature for promoting development of problem-solving skills in physics. This review article can be valuable in helping instructors develop students' problem-solving, reasoning, and metacognitive skills in physics and other related disciplines. Additionally, this review article is relevant across educational contexts in countries that may have different educational paradigms and challenges.

## I. Introduction

Research in problem-solving as well as approaches to integrate problem-solving in learning science concepts started early on in both cognitive science and education [1-12]. Understanding problem-solving has historically been associated with awareness of the research conducted by cognitive scientists. Indeed, some of the most cited advancements in the field of problem-solving have been by cognitive scientists, e.g., Chi et al., Simon and Simon, and Larkin et al., who often used physics as a medium for studying problem-solving [4, 6, 11, 13]. Furthermore, the early work of the founders of the field of problem-solving in physics, e.g., Reif, explicitly noted in their models and research design that they built on the work of cognitive scientists and even published in cognitive science journals despite having a background in physics [7]. Therefore, both historically and conceptually, understanding problem-solving in physics requires an understanding of the rich and impactful cognitive science literature on problem-solving. For this reason, in this article, we first provide a review of physics problem-solving informed by the findings of both cognitive science and physics education research before providing a concrete example of the value of effective problem-solving strategies in physics despite primarily using equations to solve problems.

Before discussing "problem-solving" research, we define what we mean by "problem-solving". We adopt the definition of Newell and Simon, and Reif [14,15] that problem-solving is a novel activity that involves devising a strategy using a sequence of steps to reach a specific goal in a limited amount of time, and in the context of physics, problems can be both conceptual or quantitative. A lot of progress has been made in learning how humans engage in these processes of reasoning through situations for which there is a gap between where the individual is and a goal they want to reach, and where they are initially unclear on how to close the gap. It has often been argued that for one to be involved in problem-solving, the activity must be novel in that one should not know beforehand the steps involved and must be actively making decisions (thinking on their feet) based on available information [14]. In the language of cognitive scientists, problem-solving is "goal directed behavior" [16], and early research to design computational models that mimic expert problem-solving showed that problem-solving is complex [1, 13, 17, 18], something that physics education researchers have generally agreed on [15, 19-25]. Moreover, from this perspective of novelty being critical, a physics "expert" (typically a physics graduate student or instructor in the physics problem-solving literature), would not engage in "problem-solving" when faced with a typical end of chapter problem because the approach is often "obvious" (automatic) to the expert from the start. Even when faced with a problem for which the solution is not immediately obvious, the vast array of experience in problem-solving an expert has results in intuition, such that the expert will have a good hunch about productive ways to approach the problem [23, 26].

To help students develop expertise in any area of physics, one must first ask how experts, in general, compare to novices in terms of their knowledge structure and their problem-solving, reasoning, and metacognitive skills. According to Sternberg [27], some of the characteristics of an expert in any field include the following: (1) having a large and well organized knowledge structure about the domain; (2) spending significant amount of time in determining how to represent problems before searching for a problem strategy (i.e., analyzing the problem and planning the solution); (3) developing representations of different problems based on deep underlying structural similarities between problems; (4) working forward from the given information in the problem and implementing strategies to find the unknowns; (5) efficient problem-solving—when under time constraints, experts solve problems correctly faster than novices; and (6) accurately predicting the difficulty in solving a problem. Additionally, experts are more flexible than novices in their planning and actions [28-30]. Individuals' expertise in a domain can span a wide spectrum on a continuum [9, 15]. With this caveat in mind, here we refer to physics instructors or graduate students/teaching assistants as experts and students who are learning to be physics experts as novices.

Experts have more robust metacognitive skills than novices. Metacognitive skills or self-regulatory skills, refer to a set of activities that can help individuals control their learning [31-34]. The three main metacognitive skills are planning, monitoring, and evaluation. Planning involves selecting appropriate strategies to use before beginning a task. Monitoring is the awareness of comprehension in light of the problem and task performance. Evaluation involves appraising the product of the task and reevaluating conclusions. Metacognitive skills are especially important for problem-solving and learning in knowledge-rich domains such as physics. For example, in physics, students benefit from approaching a problem in a systematic way, such as analyzing the problem (e.g., drawing a diagram, listing knowns and unknowns, and predicting qualitative features of the solution that can be checked later), planning (e.g., selecting pertinent principles or concepts to solve the problem), and evaluating (e.g., checking that the steps are valid and that the answer makes sense) [15, 18, 35-45].

If our goal is to help students become proficient problem solvers in physics, whether at the introductory or advanced level, we must also contemplate whether there is something special about the nature of expertise in physics over and above what we know in general about expertise, e.g., what is needed for becoming an expert tennis or chess player or music performer. Physics is a discipline that focuses on unraveling the underlying mechanisms of new physical phenomena in our universe. Physicists make and refine models to test and explain physical phenomena that are observed or to predict those that have not been observed so far. A cohesive physical model requires synthesis of both conceptual and quantitative knowledge. Therefore, an important aspect of expertise in physics is the ability to make appropriate connections between physics concepts necessary to understand physical phenomena and relevant mathematics [15]. Indeed, in physics, there are very few fundamental laws which are encapsulated in compact mathematical forms and learning to unpack them can help one develop expertise in problem-solving and organize one's knowledge hierarchically. In particular, developing expertise in physics entails making appropriate math-physics connections in order to meaningfully unpack, interpret and apply the laws of physics and use this sense-making process to develop a good knowledge structure of physics and solve novel problems in diverse situations. It is important to recognize that meaningful sense-making to unpack, interpret and apply the laws of physics, develop and organize one's knowledge structure, and retrieve relevant knowledge to solve complex physics problems is an iterative dynamic process. Appropriate reflection and metacognition during problem-solving is required to give individuals an opportunity to refine, repair and extend their knowledge structure and propel them towards a higher level of expertise.

It is also important to recognize that experts' intuition about problems can result in what is commonly described as the "expert blind spot" [46]. This implies that it is often difficult for experts to put themselves in students' shoes and evaluate the difficulty of a problem from students' perspective. This blind spot can make it difficult for them to be able to scaffold student learning and help them develop expertise in problem-solving [23]. In particular, when experts repeatedly practice problems in their domain of expertise, problem-solving and related metacognitive skills may become automatic and subconscious [31-33]. Therefore, unless experts are given a "novel" problem, they may go through the problem-solving process in an automated manner without making a conscious effort to plan, monitor, or evaluate their work. Thus, experts may have internalized certain processes involved in problem-solving to such an extent that they may not even be aware of some of them [15] and thus have difficulty providing appropriate scaffolding support to help students develop problem-solving skills.

Furthermore, since students do not have similar intuitive knowledge, they often have difficulty even knowing what considerations are productive when making problem-solving decisions (e.g., the common student experience of "not knowing where to start"). If experts are not aware of the processes involved in problem-solving in order to help students (because those processes have become automatic), they would benefit from asking students to think aloud while solving problems. Then, performing a cognitive task analysis [47] of the process of problem-solving by students can help one recognize the expert blind spots and provide a roadmap for helping students who need scaffolding support in interpreting and describing knowledge and planning the solution [15, 48, 49]. This is why research on problem-solving and effective approaches to fostering problem-solving become important for physics instructors, i.e., it can provide both a theoretical understanding of the main issues involved in problem-solving as well as practical tools to use in the classroom.

The ability to solve problems effectively within a domain requires deliberate practice to develop [50]. Deliberate practice entails being metacognitive and analyzing ones' performance in real time and making deliberate choices to improve performance. This process can be aided greatly by scaffolding via a cognitive apprenticeship model, i.e., modeling the criteria of good problem-solving, providing students with opportunities to practice effective problemsolving strategies in diverse problem contexts while receiving prompt feedback and support [51], and finally reducing the feedback gradually to help students develop self-reliance. Moreover, using this approach, problem-solving can be used to help students develop a robust knowledge structure and be able to transfer their learning from one situation to another [52-54]. For example, with appropriate scaffolding support, students can learn to recognize that all Newton's laws problems have the same underlying principle involved even though they may have very different surface features.

## II. Issues central to fostering effective problem-solving

Since problem-solving in physics necessarily involves understanding of physics concepts and being able to apply them appropriately to solve problems involving different physics principles in a variety of contexts, we must recognize that the interpretation, description, and application of these concepts is complex [24, 55-58]. In particular, failure can happen at different points in the problem-solving process, e.g., due to not having the knowledge, or having the knowledge but not recognizing that it is useful and not retrieving the knowledge, or retrieving it but applying it incorrectly. **Making decisions during problem-solving requires processing information in working memory** [4, 14, 16, 59, 60], and people's ability to process information in working memory is limited and directly related to their reasoning ability [61, 62]. Therefore, helping students become good problem solvers in a domain requires an understanding of the limited information processing capacity of students who are learning to become experts. This means that consideration of students' initial knowledge and how to provide appropriate scaffolding while minimizing their **cognitive load** during problem-solving is an important aspect of fostering effective problem-solving approaches [10, 59, 63-65]. In particular, to help students develop expertise in problem-solving, it is important for instructors to be familiar with students' relevant initial knowledge and help them manage their cognitive load by providing appropriate scaffolding support. **Expertise in problem-solving and the ability to effectively solve physics problems are also related to how knowledge is organized** and how easily a particular knowledge organization may facilitate retrieval of concepts and procedures relevant for problem-solving [3, 6, 7, 66-69]. The well-organized knowledge of experts in a domain such as physics is chunked (different knowledge pieces are connected to each other in chunks and can be accessed together) which helps reduce cognitive load while solving problems. Additionally, as noted, students must make decisions when problem-solving to reach specific goals, so **problem-solving requires continuous monitoring of one's thinking and using effective heuristics and strategies** designed to facilitate reaching those specific goals. In particular, being able to make judicious decisions when problem-solving and engaging in self-monitoring of progress as well as using feedback to change the course while solving problems require problem solvers to engage in **metacognition**, which as noted earlier, refers to people's ability to be reflective about their own thinking.

In short, prior research has found the following three factors to be pivotal in effective problem-solving:

- Information processing and cognitive load
- Knowledge organization
- Metacognition

### A. Information processing and cognitive load

This research is often conducted by cognitive scientists and involves both physics and non-physics contexts. It is important not only for understanding problem-solving in general and how to help students become good problem solvers, but also in understanding the impact of psychological issues such as math and physics anxiety, stereotype threat, and lack of social belonging since these have all been linked in some ways to information processing in working memory [70-72]. These issues often negatively impact problem-solving performance.

The cognitive revolution started with Miller's famous work on the magical number $7\pm2$ [61], which jumpstarted the analysis of problem-solving behavior using a cognitive lens. Research in cognitive science on problem-solving with regard to information processing generally argues for two broad components of human cognition [59]: long term memory (LTM) and working memory (WM). Working memory has a finite holding capacity while processing information of roughly 7 "slots" [61] while LTM does not appear to have any limits in the amount of information it can store. Kyllonen and Christal and others found [62, 73] that reasoning ability is determined by the size of the chunks in the working memory. Thus, during the problem-solving process, working memory receives information both from LTM as well as from sensory buffers (e.g., eyes, ears, hands), and this information is processed in real time according to rules selected by the problem solver. Since the amount of information that can be processed at any given time in the working memory is finite, one must first recognize what relevant information from LTM should be retrieved for the purposes of problem solving as well as carefully select the necessary information needed to be processed at a particular time to move forward with the solution.

Sweller found [10] that without guidance and support, many students employ means-end analysis when solving problems, which broadly speaking entails continuously considering the target (goal of the problem) as well as the current state (e.g., what is known) and viewing problem-solving as a process of figuring out what to do to reduce the distance between the target and the current state without big-picture considerations or use of an overarching strategy. This is contrasted with experts who often tend to work forward [16, 55, 74, 75] e.g., by starting with a qualitative analysis of the problem which involves use of multiple representations that make further analysis easier. They then plan the solution before implementing the plan. Sweller pointed out that students' means-end analysis is more resource intensive because means-end analysis entails using some of the WM slots to process the end goal and possible approaches to get there. This reduces processing capacity for what is relevant to process at a given time in order to move forward with a solution. Additionally, due to lack of a systematic approach, students are often using all of their WM slots to process information related to getting closer to the goal, and they are unlikely to have cognitive resources left for engaging in metacognition when using means-end analysis. Sweller notes, e.g., that "Conventional problem-solving activity via means-ends analysis normally leads to problem solution, not to schema acquisition." [10].

Regarding how much information can be stored in working memory, research has found that experts use chunking to extend the limits of their working memory. A famous study by Chase and Simon [2], found chess experts to be able to keep significantly more information in their working memory than novices due to what they referred to as "chunking". In the experiment, participants were shown a chess board with chess pieces on it for a very short time. The chess board would be taken away and the participants were asked to reconstruct the board. The chess experts were able to remember significantly more than just $7\pm2$ pieces when the board they were asked to reconstruct was the middle of an actual chess game, but when the pieces were randomly placed on the board, they could only recall the usual $7\pm2$. Chase and Simon kept track of <u>*how*</u> the experts reconstructed the chess board and concluded that it was not proximity that related to how they grouped different pieces together but rather relationships of attack or defense. The experts grouped pieces together in "larger perceptual chunks" based on a larger sub-configuration of figures. In other words, experts have extended the limits of their working memory capacity with chunking because they can incorporate a larger number of individual items in their working memory due to the relationships between those items even though the number of available "slots" has not changed.

Similarly, when engaged in problem-solving, physics experts often group several pieces of information together into a single chunk which would take up one slot in working memory; however, for a physics novice (e.g., a beginning student striving to develop expertise) those pieces of information could seem disparate and require different slots in order to be processed. For example, an expert could group together information about a vector such as its magnitude, direction, and *x* and *y* components into one single memory slot because of the relationships that connect them. In contrast, a novice could perceive these pieces of information as distinct and require one slot of working memory for processing each. Thus, the amount of information that a novice can process at any given time while engaged in problem-solving is reduced compared to an expert, because experts can chunk information into one single slot, whereas novices typically cannot. The amount of information that must be processed at any given time while engaged in problem-solving in order to make progress on a solution is known as cognitive load. Due to their reduced information processing capabilities, introductory students can experience cognitive overload when solving problems because the amount of information that must be processed overloads the processing capacity of the working memory.

Sweller developed cognitive load theory in an effort to explain how people learn and extend their knowledge [10]. Cognitive load theory is based on a view that the knowledge structures stored in LTM are combinations of elements, known as schemas [10, 60, 64] which, although not known precisely, can be discerned through experimental research. According to Sweller [10], learning requires a change in the schemas stored in LTM because the main difference between experts and novices in a domain is that experts possess those hierarchical schemas, while novices do not. As learners progress from novice to expert, their performance on problem-solving tasks specific to the domain learned increases because the cognitive characteristics inherent in processing the material are altered so that the material can be processed more efficiently. This suggests that as learners progress in their problem-solving abilities, they learn to group various pieces of information which are often used together in solving specific problems (i.e., chunking, as described earlier), for example, the components of a vector when dealing with problems involving vectors. This in turn means that their ability to process the necessary information when engaged in problem solving increases. Support for this idea is provided by prior research which has found that as the expertise of an individual increases in a particular field, their cognitive load decreases [60]. Sweller argues that, since information is first processed in working memory which has a finite processing capacity, in order for a learner to acquire the desired hierarchical schemas, instructional strategies must be designed to reduce cognitive load. It is therefore not surprising that many instructional strategies developed by physics education researchers, although not necessarily based on Sweller's cognitive load theory, attempt to reduce cognitive load.

Also, helping students become proficient problem solvers in the physics classroom can be enhanced by students interacting with their peers [76, 77]. Within the cognitive perspective discussed here, collaborative problem-solving can reduce cognitive load and help students solve problems successfully as well as use problem-solving as a learning opportunity. Additionally, collaborative learning can help reduce the cognitive load via distributed cognition [78-80]: the fact that the working memory of collaborators is being pooled while they solve problems together means there is more storage for information processing. This phenomenon may at least partly explain why when two students work together, they co-construct knowledge and are able to solve problems correctly that neither student could solve individually roughly 30% of the time [81]. Therefore, instructors can take advantage of collaborative group problem-solving [76, 77, 82] in which positive interdependence and individual accountability (e.g., following the group work by an individual quiz) are both incorporated. Prior research on physics group problem-solving suggests [77] that when students are not assigned specific roles, groups of three are optimal, using mixed-ability groups is more beneficial than using similar ability groups (i.e., groups of students with similar ability). Also, in mixed-gender groups, those with two female students and one male students exhibit more productive group dynamics and collaboration compared to groups with two male students and one female student.

Additionally, emphasizing explicit problem-solving heuristics and ensuring that students routinely use these heuristics when solving problems can also help reduce their cognitive load during problem-solving. This reduction in cognitive load results in freeing some WM slots for processing other information, e.g., recognizing the deep underlying structure of the problems, i.e., "seeing the forest from the trees", or engaging in metacognition. Using appropriate representations can facilitate the problem-solving process and is an integral part of effective problem-solving strategies of working forward rather than using means-end analysis. Additionally, since math is the language of physics, mathematical procedures used while solving physics problems can incur significant cognitive load for many students. Some have argued for giving students problems stripped of mathematics to get them to reason conceptually first and only incorporating the complexities of mathematics later or asking students to just carry out the qualitative analysis and planning stages of multiple problems, an approach that reduces cognitive load [83]. For example, in the context of teaching physics problem-solving, Van Heuvelen has advocated for first exploring concepts using diagrammatic and graphical representations and suggested that only after gaining sufficient mastery working with those representations, students should explore the concepts mathematically [56]. In particular, being aware of these cognitive load issues in learning and problem-solving is at least partly what informed Van Heuvelen's Active Learning Problem Sheets (ALPS) in which students started exploring a new concept using a variety of diagrammatic and graphical representations before adding the complexities of mathematics. The ALPS approach recognizes the limited processing capacity of students due to the constraints on the working memory as described by the cognitive load literature [10, 60, 84]. The design of the ALPS approach helps reduce students' cognitive load during the initial stages of learning new concepts as well as helps students make connections and build a better knowledge structure of the concepts learned.

Indeed, research suggests that a good diagram encodes information in a manner more efficient for processing the information during problem-solving [5]. A good diagram can also distribute cognition and serve as an external storage such that information doesn't have to be stored in an individual's WM memory, which can help extend the limits of information processing during problem-solving [78]. Distributed cognition and reduction in cognitive load have also been used in other studies to explain (at least partly) why diagrammatic representations facilitate information processing during problem-solving and why drawing a diagram is a representational task that students should be taught explicitly to develop expertise in physics problem-solving.

### B. Knowledge organization

The research investigating the impact of knowledge organization on problem solving was often conducted by cognitive scientists and involved both physics and non-physics contexts. The importance of knowledge organization in developing expertise in problem-solving was recognized early on and then researchers moved on to focusing on specific strategies to help students organize and extend their knowledge structure. As a consequence, most studies cited here are relatively early in the history of problem-solving research.

Early studies of expert-novice differences often focus on how knowledge is organized [6, 8, 9] based on, e.g., how knowledge is retrieved during problem-solving in a think-aloud protocol. Researchers sometimes used think-aloud interviews to infer what the knowledge structure of experts and novices looked like and represent it schematically. For example, in their classic study, Chi et al., [6] found that novices categorize problems based on surface features while experts categorize them based on deep features related to problem solution, and more importantly, how this categorization approach is based on the way knowledge is organized: experts tend to organize their knowledge hierarchically with core concepts at the top and ancillary and less important concepts towards the bottom, contrasted with novices who tend to organize their knowledge around facts and formulas that are only loosely connected. Similar findings have been reported in mathematics [85]. More recent research has used the categorization task in large enrollment introductory courses [86, 87] and found that students' ability to categorize the problems fell in a continuum with some students categorizing the problem very similarly to experts, others much closer to novices, and everything in between. The study has also been replicated with students in an upper-level quantum mechanics course [88]. The hierarchical knowledge organization of experts facilitates recognizing the deep features of a problem because it is organized around principles at the top level followed by more ancillary concepts and it is also efficient at producing the correct approach for solving a problem. During novice problem-solving, often it is not that the knowledge is necessarily missing in the student's knowledge structure, but rather how accessible the knowledge is to searches based upon the cues in the problem statement (which depends on how well-organized the knowledge structure is around fundamental principles and concepts). Some have suggested that well-constructed categorization tasks could be used as instructional tools to encourage students to organize their knowledge around fundamental principles [89, 90], and others have suggested that such tasks could be useful for assessment [35, 87, 91].

Another cornerstone finding with regard to the importance of knowledge organization is found in Eylon and Reif's 1984 work on knowledge organization and problem-solving [7]. In the study, students learned the same content with knowledge presented either hierarchically or sequentially. Students then worked on both "complex" and "local" acquisition tasks and the researchers found that students performed significantly better on the complex tasks when taught the content with knowledge presented hierarchically. The researchers concluded that having a hierarchically-organized knowledge structure facilitates performance on complex tasks that require processing an appreciable amount of information, which is precisely what problem solving is. Similarly, others [3] have argued that just having the knowledge is not enough, in order for that knowledge to be retrieved appropriately during problem solving, it must also be organized in a useful manner.

We should stress that having a knowledge organization which facilitates problem-solving and having good conceptual understanding are strongly intertwined. As stated earlier, experts' knowledge is organized around key principles and concepts with those that are foundational at the top. Someone with this type of knowledge organization would also be someone who has a good conceptual understanding. A good knowledge organization/good conceptual understanding can help learners recognize the underlying structure of a problem so that they are not distracted by the surface features of a problem as in the classic Chi et al. study of problem categorization. This in turn facilitates transfer of learning which requires the recognition that many problems can be solved with using a similar underlying approach, even if the specific details when solving the problems can be very different.

Furthermore, it's important to be aware that in physics, students don't start out as "blank slates"; rather, their incoming knowledge often consists of many preconceptions they have developed from observing how the world around them works [92]. These preconceptions are often naïve interpretations of the world which can contain many incorrect notions and they influence student learning and problem solving. This suggests that in order for students to develop the kinds of knowledge organization that facilitate effective learning and problem-solving, they must undergo a conceptual change [93], which makes the learning of physics more challenging. The research on students' pre-conceptions and framework of conceptual change suggests that creating a cognitive conflict and putting students in a state of "disequilibrium" [94] is important for helping them learn physics. Once students are in a state of disequilibrium, they can be scaffolded and guided to recognize that their prior notions are insufficient or inadequate to explain observed phenomena, which can help students repair and reorganize their knowledge structure in a manner conducive to effective problem-solving.

A great example of incorporating knowledge organization for problem-solving in instructional design is Bagno et al.'s approach [66, 67], which "integrates problem-solving, conceptual understanding, and the construction of a knowledge structure". One of the key characteristics of the approach is that students are led to construct their knowledge structure through problem-solving and generating concept maps that link the central concepts of the topics. Students solve problems, reflect on the central concepts that relate the problems and represent them in different forms, develop and elaborate the concepts to overcome conceptual difficulties, then apply this knowledge to non-familiar problem-solving, and lastly, they link the new concept to their prior knowledge [66]. Additionally, students benefit from making explicit links (via concept maps) between the concepts of mechanics and electromagnetism [67]. The goal of this instructional design is to 1) support students in organizing their knowledge around key concepts and 2) make use of their knowledge organization when solving problems. Similarly, Van Heuevelen's ALPS approach includes explicit activities designed to help students organize the concepts learned in a useful manner as well as link them to previous units [83].

The Hierarchical Analysis Tool (HAT) is another example [48]. The HAT is a computer program which is designed such that it takes students through a hierarchical approach when solving problems: students answer a series of questions which start with broad principles and gradually get more specific, with subsequent questions depending on the choices made. After the student makes all the appropriate choices, the HAT framework comes out with a system of equations which are consistent with the choices made by the student. If the student made the correct choices, the equations accurately describe the system and can be used to solve the problem. The researchers found that students who used the HAT tool exhibited improved problem-solving performance compared to students who solved problems traditionally, and they conclude that "findings indicate that performing this type of qualitative analysis for a relatively short period of time results in statistically significant shifts toward problem-solving behavior observed in experts." This positive outcome is also related to metacognition involved in different stages including performing a qualitative analysis and planning the solution while answering the questions. Maloney [75] provides an excellent overview of many of these early studies on knowledge organization.

### C. Metacognition

As noted, metacognition during problem-solving refers to being reflective about one's own thinking; being able to assess accurately where one currently is in their problem-solving process, being able to devise strategies to improve their current state and also reflecting back upon the problem-solving process to learn from it. The early research on problem-solving, e.g., in chess [2, 95], in mathematics [84, 96, 97], or in physics [9, 13, 17, 55, 74] focused explicitly on learning how experts solve problems and using this knowledge to design instruction to help students learn effective problem-solving strategies. The research on expert-novice differences in problem-solving [6, 13, 55, 98, 99] found that experts engage in metacognition throughout problem-solving, e.g., they spend significant time and effort analyzing, interpreting and describing the problem and converting it into a representation that facilitates progress before planning the solution, implementing the plan, then checking their solution and reflecting upon what they learned from the problem-solving process. Metacognition at all stages of problem-solving, especially during the initial qualitative analysis stage, can be very productive in solving the problem successfully. Moreover, while solving problems, experts also engage in metacognition to sharpen the necessary procedural knowledge and unpack the physical laws used to solve the problem efficiently. Beginning students must be given explicit opportunities and guidance to engage in metacognition while problem-solving.

Chi et al.'s categorization study [6] is a metacognitive task which asked experts and novices to group problems based on their solutions. Their findings indicate that experts are more likely to group problems based upon the underlying physics principles involved and not get distracted by surface features. In order for students to develop expertise in problem-solving, some of the earliest attempts to teach problem-solving (e.g., Polya's well known book "How to Solve It") emphasized the importance of being systematic and analyzing problems qualitatively before choosing a solution approach [74]. Research studies since then [20, 36, 41, 49, 52, 53, 75-77, 91, 100-115] have provided a more nuanced picture of exactly what types of problem-solving heuristics and strategies may work in what contexts and different researchers have developed different approaches and programs, e.g., "GOAL" directed problem-solving by Beichner [57], or the Personal Assistant for Learning or PAL [98], or computer coaches [113], etc. Nevertheless, the general framework of qualitative analysis and planning that involves significant use of representations followed by an implementation stage, then an evaluation stage and a broader reflection on what was learned from the problem-solving process has generally remained the same.

Metacognition can be useful at all stages of problem-solving. In the qualitative analysis stage of problem-solving, it is important to both help students realize the importance of using multiple representations, as well as help them learn how to draw useful diagrams and use them appropriately [9, 15, 36-39, 42, 43, 83, 105, 106, 116-120]. In physics, what a useful representation is depends on the context, and there are often rules about both constructing useful diagrams as well as using them. In introductory mechanics for example, for problems involving Newton's laws, Reif suggested that a useful representation would separate the system from the environment, represent its motion and identify interactions, e.g., non-contact forces are identified first, and contact forces are identified from contact points [15]. He also illustrated these ideas with specific examples, e.g., students can first draw a diagram including information from the problem description, and use it in specific ways (e.g., mark contact points with the names of the action/reaction forces) that help them generate the useful system description which identifies the forces acting on each object of interest. This system description in turn is used to generate the equations that describe the physics in the problem which will then be used to solve for the target variables.

Heller and Reif [9] investigated the impact of being systematic and metacognitive during physics problem-solving in which students were either 1) shown effective problem-solving strategies and required to use these strategies in a problem-solving session, 2) shown effective problem-solving strategies, but allowed to use any strategies they wanted during a problem-solving session, or 3) not shown any information about effective problem-solving strategies. They found that students from groups 3, 2, and 1 performed progressively better on a subsequent problem-solving task. Additionally, students themselves were sometimes amazed at being able to solve the more challenging problems. They recognized that the problems were challenging and didn't necessarily believe that they could solve them, but the systematic strategies they learned and employed led them to solve the problems successfully. This type of experience also builds self-efficacy or belief in one's ability to solve problems [121], which helps motivate students to further engage meaningfully with problem-solving [121-123].

In mathematics, to estimate the extent of metacognition at each stage of problem-solving, Schoenfeld [32] encoded the problem-solving strategies used by students in graphs which provide a description of how much time students spent reading, analyzing, exploring, planning, implementing, and verifying. He found that beginning students tend to spend a lot of time exploring, for example, making a guess and then spending significant time attempting to prove that the guess was correct, but often not stopping to ask themselves if they are making progress or if their guess was reasonable. Without training, students tended to spend little time analyzing or planning the solution to a problem. This lack of planning and metacognition [31] is detrimental to successfully solving the problem as well as learning from problem-solving.

Metacognition is also critical for learning from problem-solving because students should use problem-solving as an opportunity to think about what they have learned more broadly and how it can help them repair, organize, and extend their knowledge structure. Why were certain principles useful to solve the problem correctly but others were not? If I change certain parameters in the problem, how will that impact the results? What do we learn from this problem that can be applied to other problems? What are the features of the problems that will help me recognize that this same principle of physics is applicable in those situations? Metacognition after having solved the problem is an important part of problem-solving and asking students to diagnose their mistakes after solving problems by providing them incentives to do so can be productive in helping students learn physics [124-127]. In order to help students become adaptive experts [29-30], they must be given opportunities and incentives to use problem-solving as an opportunity to recognize the deep features of the problems so that they can transfer their learning to other novel problems using the same underlying principles but very different contexts. In other words, developing expertise in physics requires using problem-solving as a learning opportunity and becoming proficient in using a handful of laws of physics to solve a wide range of problems.

### III. Incorporating problem-solving into instructional design

#### A. Cognitive apprenticeship model

As briefly mentioned earlier, students can learn effective problem-solving approaches via cognitive apprenticeship [51], a broad overarching model for effective teaching and learning. In the context of problem-solving, cognitive apprenticeship includes modeling effective approaches to problem-solving, providing coaching, scaffolding and prompt feedback as students practice problem-solving, and gradually fading the support in order to help them develop self-reliance. Modeling is important because students first need to observe expert performance in solving problems, which involves all stages of problem-solving including conceptual analysis, planning, evaluating the solution, and reflecting upon what was learned. Without guidance, students often solve problems in counterproductive ways (e.g., using means-end analysis). Furthermore, students must explicitly be provided opportunities to practice effective problem-solving strategies and receive prompt feedback and scaffolding with a gradual weaning of support as students become self-reliant. For example, in Van Heuvelen's Overview: Case Study Physics (OCS), the cognitive apprenticeship model was used as a guide [83]. It is particularly important to develop students' facility with representations and use them in subsequent problem-solving. In the OCS approach, students first explored new concepts using various representations and, in this process, developed facility with describing the concepts visually. Moreover, the representational facility is first developed outside of larger problem-solving tasks in the OCS approach. This is especially useful for learning physics problem-solving, and students are required to describe the various aspects of the problems they solve later by using strategies learned at the beginning of the unit in the initial stage of problem-solving. Others have developed interactive problem-solving tutorials inspired by the cognitive apprenticeship framework [45, 128-131].

#### B. The ICQUIP framework

The earlier discussion on metacognition suggests that solving physics problems productively requires use of both conceptual and quantitative knowledge and as we have seen, prior research on problem-solving often emphasizes the importance of doing a qualitative (i.e., conceptual) analysis as the first step in solving a problem regardless of the type of problem solved [9, 15, 57, 76, 101, 105, 113]. Based on this prior research, a framework for teaching problem-solving in physics called "Integrating Quantitative and Conceptual Understanding in Physics" [132] or ICQUIP has been proposed. The main idea behind this framework is that students require significant practice in interpreting symbolic equations and drawing qualitative inferences from them in order to use problem solving as an opportunity to develop a functional understanding of physics. If this integration of conceptual and quantitative aspects of physics is lacking in a course, students may view problem-solving as merely mathematical, or in other words, believe that solving problems is all about manipulating equations and consequently view using productive representations to make sense of how the concepts are applied to solving problems as unnecessary. Prior research shows that without guidance and support (as well as incentives), students often solve problems by focusing on superficial features and often rely on pattern matching with previously solved problems or example solutions. It is also important to point out that traditional instruction may *reward* this type of problem solving when it does not emphasize and incorporate conceptual aspects into problem-solving as well. Some researchers have found [133] that interactive engagement courses which primarily focus student group work around conceptual questions do not necessarily result in improved problem-solving performance, thus highlighting the importance of integrating both conceptual and quantitative problem-solving when engaging students in interactive activities.

We emphasize that the ICQUIP framework should be integrated within the broader cognitive apprenticeship model. In other words, modeling should be used to show students how to combine conceptual and quantitative knowledge, then students engage in extended practice while receiving guidance and support (i.e., the coaching and scaffolding stage), and lastly weaning should be used to help develop self-reliance. For example, recitations are ideal for the coaching and scaffolding stage where students can solve problems in small groups and the scaffolding is reduced (or removed) when students work individually on homework.

An approach described in the literature consistent with the ICQUIP framework involves using quantitative problems followed by either isomorphic or relatively similar conceptual problems [134]. This approach is most effective when using conceptual problems that probe common student alternate conceptions because they can help students recognize how to use the concepts learned in physics to reason about situations in which they would otherwise be very likely to answer incorrectly because they rely their "gut feeling". For example, students in a large class were divided into two groups: one group solved only a conceptual problem in which there was a strong alternate conception and another group first solved a quantitative problem before solving the conceptual problem. The quantitative problem required students to perform a calculation that was useful for the conceptual problem. Even without explicit guidance that the two problems are related, students who solved both were significantly less likely to solve the qualitative problem by using their "gut feeling" (roughly half the students solved the qualitative problem correctly compared to only 16% in the group that only solved one problem). Discussions with students suggested that even without explicit guidance, students recognized the similarity between the two problems and made use of what they learned from solving the quantitative problem when solving the qualitative one.

Another example comes from Mazur [135] who restructured his course and included many conceptual questions during lectures, and also emphasized conceptual reasoning on examinations. He found that both student conceptual understanding as well as problem-solving performance improved, although, a larger improvement was observed in conceptual understanding. Similarly, McDermott [136] used traditional quantitative problem-solving activities in recitation with conceptual tutorials and found improved student performance on conceptual problems, but also at least as well or better performance in solving quantitative problems.

### C. Research-based approaches to foster effective problem-solving

Physics problem-solving goes hand in hand with learning and organizing one's knowledge structure because problem-solving is an essential tool to help students learn and develop a solid grasp of physics concepts. For problem-solving to be effective, instructors must be aware of students' prior knowledge [137, 138] and design problem-solving tasks that are just beyond students' current capability, but in which they can learn and develop expertise with appropriate support and guidance. This is also the idea behind Piaget's "optimal mismatch" [94] and Vygotsky's "zone of proximal development" [139]. We note that engaging in these types of tasks while practicing effective problem-solving strategies is precisely what deliberate practice is [50]. It is also important for instructors to reward and incentivize students for using effective problem-solving strategies since assessment drives learning, as well as use research-based approaches to assessment of problem-solving using a good rubric [140].

With regard to knowledge organization, we already discussed a few research-based approaches [48, 65, 66, 83]. Another example comes from work by Scott and Reif [98] based on Reif's extensive research in problem-solving [7-9, 15, 55, 141]. The Personal Assistant for Learning or PAL is a computer program [49, 98] that made use of the reciprocal teaching method [142] to help students become proficient problem solvers. The PAL coaches students through using problem-solving strategies and then the role is reversed and students make decisions that are assessed by the computer. Reif and Scott had students in three groups, all working on a homework assignment. One group used the PAL, another used experienced human tutors, and the last solved the problems without any guidance. The groups were all matched for prior performance, and on a subsequent problem-solving task on the same content, students in the PAL and human tutor group performed similarly and significantly better than the control group.

This early PAL computer program has undergone many updates and revisions, and currently the University of Minnesota physics education research group maintains existing computer coaches based on the original PAL computer program as well as creates new ones by modifying the features based upon research [113]. The computer coaches, which now use more sophisticated methods than in the late 90s are designed to "help students develop metacognitive problem-solving skills" [113] and the cognitive apprenticeship model is included in the design of these coaches. The authors recommend using the computer coaches as part of homework, with the coaches being required so that students first go through expert-like decision making while receiving feedback and guidance from a coach and then they can practice on their own. When used in class, students overwhelmingly recognize the benefit of the coaches in teaching them how to solve problems, more so than their homework system, in-class lectures, or textbook. Others have used similar considerations when creating other interactive online tutorials designed to foster physics problem-solving [54, 131].

One issue with any tool students use on their own is the extent to which they engage meaningfully with it. It's not just the time students spend on a particular task that matters for learning from problem-solving, but what is critical is that they are mentally engaged. For example, prior research [45, 129, 130] found that when students studied on their own from guided problem-solving tutorials with context-rich problems (which guided students to make decisions in a fashion somewhat similar to the computer coaches), many students simply glossed over them instead of thinking through each sub-problem and thus derived little benefit from them. But when students worked through the problem-solving tutorials in the presence of a researcher who only instructed students to think aloud while solving the problems and ensured that students were using the tutorial properly and thinking through the hints and guidance as directed, they did exceedingly well on subsequent transfer problems, with performance nearing the ceiling. This finding is somewhat similar to Heller and Reif's study [9] in which students who were required to use effective problem-solving strategies significantly outperformed students who were only provided with those strategies.

Another strategy for teaching problem-solving is to have students diagnose their own mistakes [125-127, 143]. In the studies by Yerushalmi et al. [125-126], the intervention that best helped students learn from their mistakes and improve their problem-solving incentivized them to correct their mistakes. They were required to use their book and lecture notes to correct their mistakes for 50% points back on the problems before they were provided the correct solutions. In other words, students in this intervention group had access to solutions in textbook problems or problems in their class notes for diagnosing their own mistakes. They had to recognize the connection between their own solution and any similar textbook solution or solution in their notes. They then had to struggle to self-diagnose their mistakes. It is likely that these students experienced more cognitive involvement in their self-diagnostic activity than students in other intervention groups who were provided more support (e.g., a detailed solution) for self-diagnosis. The authors noted [126] that "the more the external support, the less we could differentiate between students whose self-diagnosis was or was not accompanied by self-repair." In other words, if students are provided a lot of external support, they may perform the reflection task superficially without engaging with the underlying physics concepts deeply, which can reduce learning. There is also evidence that without this additional incentive, many advanced students may also not learn from their mistakes even if the correct solution is provided to them with their graded problem solution [127]. Similarly, many students aren't necessarily motivated to use effective problem-solving strategies (e.g., qualitative analysis, planning the solution and converting the problem into representations that can facilitate the problem-solving process) without an incentive, but if they are provided incentive, e.g., via grades, to draw diagrams themselves while solving physics problems, they often benefit from the activity [37].

Using analogical reasoning can also be a useful approach for helping students learn physics problem-solving [107, 108, 144-148]. For example, in problem situations in which there are strong alternate conceptions, analogical reasoning can be used to help students recognize how the underlying structure of the two problems (with and without alternative conceptions) are similar despite differences in surface features [107]. Asking students to solve isomorphic problems, i.e., problems that have different surface features but can be mapped onto each other due to their underlying similarity, can be a productive approach to helping students learn effective problem-solving strategies [52, 53].

Many prior research studies on problem-solving can be applied directly to classroom instruction. For example, Chi et al.'s task of grouping problems based upon similarity of solution can be used with students in classes in small groups to help them recognize the underlying similarities of problems that use the same physics principles even if the surface features are very different [89, 90]. Many of the problem-solving strategies described in detail, e.g., in Heller and Reif's 1984 study [9], can be used as models to create class activities, e.g., pertaining to what constitutes a good qualitative analysis while solving a physics problem.

One commonly recommended activity in recitation classes (briefly mentioned earlier) is collaborative group problem-solving with context rich problems [76, 77]. Heller and Hollabaugh found that when students work on problem-solving using explicit strategies in mixed ability collaborative groups, they benefit more than when working alone. In Heller and Hollabaugh's approach, students follow a prescribed problem-solving approach based on Heller and Reif [9] which involves both a visualization step and a physics description step that are more specific to creating representations related to physics (e.g., free body diagrams). Students use this strategy to solve homework problems while working in groups of 3 with specific roles: Manager (designs plans for action and suggests solutions), Skeptic (questions premises and plans) and 3) Checker/Recorder (organizes and keeps track of the results of the discussion). They arrived at three as a good number of group members for group problem-solving after conducting research and considering practical considerations [76, 77]. Potentially, four roles can be used in each group, with the addition of Explainer, someone who takes the burden of explanation and summarizing. Also, the roles should rotate from week to week. They advocate for the use of context-rich problems, both because these types of problems are closer to what one may realistically encounter in the real world and also because they are better vehicles for teaching problem-solving: students are more likely to benefit from a systematic approach when the problem is challenging. For these types of more challenging problems, students can benefit from a variety of perspectives, especially if they are being systematic with someone proposing solutions, someone being skeptical and forcing the solutions to be better articulated and someone keeping track of explaining and recording the discussion. Moreover, students having specific roles helps reduce their cognitive load because of distributed cognition [78-80] and the fact that each individual does not have to keep everything related to solving the problem in their working memory. In order for students to develop self-reliance, it is important for students to take on different roles in different weeks and keep rotating. In addition to positive interdependence in group work, it is important for there to be individual accountability [82], e.g., via individual assessment, for learning to be meaningful. Otherwise students may always require the support of others and not be able to solve problems on their own. Heller and Hollabaugh found that when using collaborative learning with context-rich problems, students' problem-solving abilities generally progress over time and students from all abilities benefit [76, 77].

Another approach to guide and support students in learning effective problem-solving strategies is to use reflection with peers [149, 150]. Students can work individually on solving specific problems, then work in groups of three to discuss which solution is the "best" one, i.e., the one that best exemplifies the use of effective problem-solving strategies. Those solutions can compete with solutions chosen by other student groups and then a whole-class discussion ensues and a winner is chosen whose solution best exemplifies effective problem-solving strategies. This type of activity can be useful in recitations associated with physics courses and has been found to result in greater use of effective problem-solving approaches, e.g., drawing diagrams, even in multiple-choice exams in which there was no grade incentive for doing so compared to a traditional recitation [150].

More broadly, recitation can be an ideal place to provide students with coaching and scaffolding in effective problem-solving strategies as well as give immediate feedback. In addition to having smaller class sizes, recitations are often taught by someone closer in age to the students, e.g., a graduate student or advanced undergraduate. This can help make the "advice" about how to solve problems, especially when couched in a practical message of being able to solve problems quickly and accurately on exams, be better received by students who, as Schoenfeld [32] remarked, have good reasons to not necessarily believe you when you tell them they are not using good problem-solving strategies. After all, they made it to college. Also, since assessment drives learning, it is important that use of effective problem-solving strategies is rewarded via grade incentives.

Regarding formative assessment more broadly, we emphasize that it is a continuous process that occurs over the entire course of instruction. When students solve problems in class or recitation working in groups, if instructors and TAs provide coaching and scaffolding about their problem-solving process, that's formative assessment. Throughout this review, we have discussed many specific tasks that could be used for assessment purposes, for example, the categorization task, concept maps, pairing conceptual and quantitative problems, the HAT tool, etc. in addition to textbook problems including those requiring application, analysis and synthesis [151-154]. Some researchers have developed rubrics specifically aimed at quantifying problem-solving performance [140, 155, 156], although, some of these are aimed more at researchers than practitioners.

## IV. Conclusion and future directions

Helping students become effective problem solvers is an important goal of many physics courses, and it is a necessary component of any instructional design aiming to help novices develop expertise. The literature on problem-solving is very broad; nevertheless, it can be broken down into three inter-related categories: information processing and cognitive load, knowledge organization, and metacognition and problem-solving heuristics. In broad terms, problem-solving requires processing information in working memory, which is limited, especially for novices who do not have sufficient experience to be able to chunk several pieces of information together into a single chunk. How students' knowledge is organized affects the strategies they use when solving problems, and in particular, the effectiveness with which they can retrieve (or not retrieve) relevant information from long term memory. To help students become effective problem solvers, they must be guided to learn and practice effective problem-solving heuristics that help them 1) work forward instead of using means-end analyses and 2) recognize the relevant information that needs to be used at each point in the solution process to facilitate moving forward with the solution. These kinds of problem-solving heuristics should help with the limited information processing capacity of students who are striving to become experts in physics, especially if they emphasize use of multiple representations to help offload some of the cognitive load and make use of distributed cognition. Group work can also help in this regard. In part 2 of this article, we will describe a research study investigating one aspect of problem-solving, namely the initial step of conceptual analysis and planning, in particular, with regards to use of diagrammatic representations.

There are several future directions that researchers can pursue. One of the findings we discuss is that students who draw diagrams perform better than students who do not draw them even if their approach primarily involves manipulation of equations [157]. While this is consistent with prior research, it's difficult to disentangle what is the cause and what is the effect, i.e., are students who are "better" at problem solving the ones who also draw more diagrams (an expert-like problem-solving strategy), or does the diagram help in some way, or both? While we have described some research on using online tools to help students learn problem-solving, it will be useful to pursue future research that provides guidance as to how to integrate these tools with an actual class, what kinds of incentives can be used [45].

Furthermore, physics education research has been increasingly investigating issues of diversity, equity, and inclusion; in particular, whether certain educational tools may have differential effects on different groups of students. For example, active learning in an inequitable learning environment may have a disproportionate positive impact on male students thus widening the gender gap on students' conceptual performance even if all groups learn more [158]. Future research could investigate the impact on various demographic groups and contexts (e.g., based upon gender, race/ethnicity, type of institution or country etc.) of various approaches that have been found effective overall for helping students develop problem-solving skills. In particular, the extent to which they help students from all demographic groups become proficient problem solvers and independent learners .

Furthermore, since the goal is always to improve students' problem-solving abilities, instructors necessarily must be part of the process. There is some research on professional development for faculty and TAs (that we have described in this review), but more research is needed to understand effective approaches for training faculty on instructional design which incorporates activities to improve students' problem-solving skills. Research on how to make instructors interested in trying effective approaches to improving student problem solving skills and what kinds of institutional support and incentives to the instructors are useful in different contexts in different countries with a variety of instructional and institutional constraints would useful. It would be useful to investigate how instructors in different countries and different types of institutions interested in improving their students' problem-solving skills use different types of problems (e.g., context-rich problems, broken-into-parts problems, traditional textbook problems, multiple-choice problems) and the impact it has on their students' problem-solving, reasoning and meta-cognitive skills under different educational paradigms, institutional constraints and culture.

## References

[1] Newell, A. et al. Elements of a theory of human problem solving. *Psychol. Rev.* **1958**, *65*, 151-166.
[2] Chase, W.; Simon, H. Perception in chess. *Cog. Psych.* **1973**, *4*, 55-81.
[3] de Jong, T.; Ferguson-Hessler, G. Cognitive structures of good and poor novice problem solvers in physics. *J. Educ. Psych.* **1986**, *78*, 279-288.
[4] Simon, H. How big is a chunk? *Science* **1974**, *183*, 482-488.
[5] Larkin, J.; Simon, H. Why a diagram is (sometimes) worth ten thousand words. *Cog. Sci.* **1987**, *11*, 65-99.
[6] Chi, M. et al. Categorization and representation of physics knowledge by experts and novices. *Cog. Sci.* **1981**, *5*, 121-151.
[7] Eylon, B.; Reif, F. Effects of knowledge organization on task performance. *Cog. Instruct.* **1984**, *1*, 5-44.

[8] Heller, J.; Reif, F. Knowledge structure and problem solving in physics. *Educ. Psych.* **1982**, *17*, 102-127.
[9] Heller, J.; Reif, F. Prescribing effective human problem-solving processes: Problem description in physics. *Cog. Instruct.* **1984**, *1*, 177-216.
[10] Sweller, J. Cognitive load during problem solving: Effects on learning. *Cog. Sci.* **1988**, *12*, 257-285.
[11] Simon, H.; Simon, D. Individual differences in solving physics problems. In *Children's thinking: What develops?* Siegler, R.S., Ed.; Lawrence Erlbaum Associates: Hillsdale, New Jersey, USA, 1978; p. 325.
[12] Bassok, M.; Holyoak, K. Interdomain transfer between isomorphic topics in algebra and physics. *J. Exp. Psych.: Learning Memory and Cognition* **1989**, *15*, 153-166.
[13] Larkin, J. et al. Expert and novice performance in solving problems. *Science* **1980**, *208*, 1335-1342.
[14] Newell, A.; Simon, H. *Human Problem Solving;* Prentice Hall: Englewood Cliffs, NJ, USA, 1972.
[15] Reif, F. Millikan lecture 1994: Understanding and teaching important scientific thought processes. *Am. J. Phys.* **1994**, *63*, 17-32.
[16] Anderson, J. Problem solving. In *Cognitive Psychology and its Implications*; Anderson, J., Ed.; Worth: New York, NY, 2000; p. 239.
[17] Larkin, J. et al. Models of competence in solving physics problems. *Cog. Sci.* **1980,** *4*, 317-345.
[18] Chi, M.; Glaser, R. Problem solving ability. In *Human Abilities: An Information Processing Approach*; Sternberg, R.J., Ed.; Freeman: New York, NY, USA, 1985; p. 227.
[19] Larkin, J. et al. FERMI: A flexible expert reasoner with multi-domain inferencing. *Cog. Sci.* **1988**, *12*. 101-138.
[20] Mestre, J. et al. Promoting skilled problem solving behavior among beginning physics students. *J. Res. Sci. Teach.* **1993**, *30*, 303-317.
[21] Holyoak, K. Problem solving. In *An Invitation to Cognitive Science*; Smith, E., Osherson, D., Eds.; The MIT Press: Cambridge, MA, 1995; p. 267.
[22] Leonard, W. et al. Using qualitative problem solving strategies to highlight the role of conceptual knowledge in solving problems. *Am. J. Phys.* **1996**, *64*, 1495-1503.
[23] Singh, C. When physical intuition fails. *Am. J. Phys.* **2002**, *70*, 1103-1109.
[24] Adams, W.; Wieman, C. Analyzing the many skills involved in solving complex physics problems. *Am. J. Phys.* **2015**, *83*, 459-467.
[25] Price, A. et al. A detailed characterization of the expert problem-solving process in science and engineering: Guidance for teaching and assessment. *CBE-Life Sci. Educ.* **2021**, *20*, 1-15.
[26] Sayer, R. et al. (). Advanced students' and faculty members' reasoning about the double slit experiment with single particles. In Proceedings of the 2020 Physics Education Research Conference, Virtual, Online, July 19-22; Wolf, S., Bennett, M., Frank, B., Eds.; 2020, pp. 460-465.
[27] Sternberg, R. Metacognition, abilities, and developing expertise: What makes an expert student? *Instruct. Sci.* **1998**, *26,* 127 (1998).
[28] Berliner, D. Expertise: The wonder of exemplary performances. In *Creating Powerful Thinking in Teachers and Students: Diverse Perspectives*; Mangerieri, J., Collins Block, C., Eds.; Holt, Rinehart, and Winston: Fort Worth, TX, USA, 1994; p. 141.
[29] Hatano, G.; Inagaki, K. Two courses of expertise. In *Child development and education in Japan*; Stevenson, H., Azuma, H., Hakuta, K., Eds.; Freeman: New York, NY, USA, 1986; p. 262.
[30] Hatano, G.; Oura, Y. Commentary: Reconceptualizing school learning using insight from expertise research. *Educ. Res.* **2003**, *32*, 26-29.
[31] Schraw, G. Promoting general metacognitive awareness. *Instruct. Sci.* **1998**, *26,* 113–125.
[32] Schoenfeld. A. (1987). What's all the fuss about metacognition. In *Cognitive Science and Mathematics Education*; Schoenfeld, A., Ed.; Lawrence Erlbaum Associates: Hillsdale, NJ, USA, 1987; p. 189.
[33] Jacobs, J.; Paris, S. Children's metacognition about reading. Issues in definition, measurement, and instruction. *Educ. Psych.* **1987**, *22,* 255-278.
[34] Morphew, J. et al. Effect of presentation style and problem-solving attempts on metacognition and learning from solution videos. *Phys. Rev. PER* **2020**, *16*, 010104.
[35] Hardiman, P. et al. The relationship between problem categorization and problem solving among experts and novices. *Mem. Cogn.* **1989**, *17*, 627–638.
[36] Dufresne, R. et al. Solving physics problems with multiple representations. *Phys. Teach.* **1997**, *35*, 270-275.
[37] Maries, A.; Singh, C. Do students benefit from drawing productive diagrams themselves while solving introductory physics problems? The case of two electrostatics problems. *Eur. J. Phys.* **2018**, *39*, 015703.
[38] Maries, A.; Singh, C. Case of two electrostatics problems: Can providing a diagram adversely impact introductory physics students' problem solving performance? *Phys. Rev. PER* **2018**, *14*, 010114.
[39] Nguyen, D.-H. et al. Facilitating students' problem solving across multiple representations in introductory mechanics. *AIP Conf. Proc.* **2010**, *1289*, 45-48.
[40] Rosengrant, D. et al.. Free-body diagrams - Necessary or sufficient. *AIP Conf. Proc.* **2005**, *790*, 177-200.
[41] Singh, C. Coupling conceptual and quantitative problems to develop student expertise in introductory physics. *AIP Conf. Proc.* **2008**, *1064*, 199-202.
[42] Van Heuvelen, A.; Zou, X. Multiple representations of work-energy processes. *Am. J. Phys.* **2001**, *69*, 184-194.
[43] van Someren, M. et al. *Learning with Multiple Representations.* Elsevier Science Inc.: New York, NY, USA, 1998.

[44] Warnakulasooriya, R. et al. Time to completion of web-based physics problems with tutoring. *Journal of the Experimental Analysis of Behavior* **2007**, *88*, 103-113.
[45] DeVore, S. et al. Challenge of engaging all students via self-paced interactive electronic learning tutorials for introductory physics. *Phys. Rev. PER* **2017**, *13*, 010127.
[46] Nathan, M. et al. Expert blind spot: When content knowledge eclipses pedagogical content knowledge. In *Proceeding of the Third International Conference on Cognitive Science;* Chen L. et al., Eds.; USTC Press: Beijing, China, 2001; p. 644.
[47] Chipman, S. et al. Introduction to cognitive task analysis. In *Cognitive Task Analysis*; Schraagen, J., Chipman, S., Shute, V., Eds.; Lawrence Erlbaum Associates: Mahwah, New Jersey, USA, 2000; p. 3.
[48] Dufresne, R. et al. Constraining novices to perform expertlike problem analyses: Effects on schema acquisition. *J. Learn. Sci.* **1992**, *2*, 307-331.
[49] Hsu, L.; Heller, K. Computer problem solving coaches. *AIP Conf. Proc.* **2005**, *790, 197-201.*
[50] Ericsson, K. et al. The role of deliberate practice in the acquisition of expert performance. *Psychol. Rev.* **1993**, *100*, 363–406.
[51] Collins, A. et al. Cognitive apprenticeship: Teaching the crafts of reading, writing and apprenticeship. In *Knowing, Learning and Instruction: Essays in Honor of Robert Glaser*; Glaser R., Resnick, L., Eds.; Lawrence Erlbaum Associates: Hillsdale, New Jersey, USA, 1989; p. 453.
[52] Lin, S.-Y.; Singh, C. Using an isomorphic problem pair to learn introductory physics: Transferring from a two-step problem to a three-step problem. *Phys. Rev. ST-PER* **2013**, *9*, 020114.
[53] Lin, S.-Y., & Singh, C. Effect of scaffolding on helping introductory physics students solve quantitative problems involving strong alternative conceptions. *Phys. Rev. ST-PER* **2013**, *11*, 020105.
[54] Whitcomb, K. et al. Improving accuracy in measuring the impact of online instruction on students' ability to transfer physics problem-solving skills. *Phys. Rev. ST-PER* **2021**, *17*, 010112.
[55] Larkin, J.; Reif, F. Understanding and teaching problem solving in physics. *Eur. J. Sci. Educ.* **1979**, *1*, 191-203.
[56] Van Heuvelen, A. Learning to think like a physicist: A review of research-based instructional strategies. *Am. J. Phys.* **1991**, *59*, 891-897.
[57] Beichner, R. Goal oriented problem solving. **1999**; retrieved from https://projects.ncsu.edu/per/archive/GOALPaper.pdf
[58] Singh, C., Maries, A., Heller, K., and Heller, P. Instructional strategies that foster effective problem-solving. In *Handbook of Research on Physics Education, Volume 1: Learning Physics*, Taşar, M.F., Heron, P., Eds.; AIP Publishing, Melville, New York, USA, 2023, pp. 17-1–17-28.
[59] Anderson, J. *Learning and Memory*. Wiley: New York, NY, USA, 1995.
[60] Sweller, J. Cognitive Load Theory. In *The Psychology of Learning and Motivation*; Mestre, J., Ross, B, Eds.; Elsevier Academic Press: San Diego, CA, USA, 2011; p. 37.
[61] Miller, G. The magical number seven, plus or minus two: Some limits on our capacity for processing information. *Psychol. Rev.* **1956**, *63*, 81-97.
[62] Kyllonen, P.; Christal, R. Reasoning ability is (little more than) working memory capacity?! *Intelligence* **1990**, *14*, 389-433.
[63] Sweller, J. et al. Cognitive architecture and instructional design. *Educ. Psych. Rev.* **1998**, *10*, 251-296.
[64] Paas, F.; Van Merriënboer, J. Variability of worked examples and transfer of geometrical problem-solving skills: A cognitive-load approach. *J. Educ. Psych.* **1994**, *86,* 122–133.
[65] Paas, F. et al. Cognitive load theory: Instructional implications of the interaction between information structures and cognitive architecture. *Instruct. Sci.* **2004**, *32,* 1–8.
[66] Bagno, E.; Eylon, B. From problem solving to a knowledge structure: An example from the domain of electromagnetism. *Am. J. Phys.* **1997**, *65*, 726-736.
[67] Bagno, E. et al. From fragmented knowledge to a knowledge structure: Linking the domains of mechanics and electromagnetism. *Am. J. Phys.* **2000**, *67*, S16-S26.
[68] Beatty, I., & Gerace, W. Probing physics students' conceptual knowledge structures through term association. *Am. J. Phys.* **2002**, *70*, 750-758.
[69] Sabella, M.S.; Redish, E.F. Knowledge organization and activation in physics problem solving. *Am. J. Phys.* **2007**, *75*, 1017-1029.
[70] Beilock, S. et al. On the causal mechanisms of stereotype threat: Can skills that don't rely heavily on working memory still be threatened? *Pers. Soc. Psych. Bull.* **2006**, *38,* 1059-1071.
[71] Beilock, S. et al. Stereotype threat and working memory: Mechanisms, alleviation, and spillover. *J. Exp. Psych.: Gen.* **2007**, *136*, 256-276.
[72] Maloney, E. et al. Anxiety and Cognition. *WIREs Cognitive Science* **2014**, *5,* 403-411.
[73] Kail, R.; Salthouse, T. Processing speed as a mental capacity. *Acta Psychologica* **1994**, *86*, 199-225.
[74] Larkin, J. Skilled problem solving in physics: A hierarchical planning model. *J. Struct. Learn.* **1980**, *6*, 271-297.
[75] Maloney, D. Research on Problem Solving: Physics. In *Handbook of Research on Science Teaching and Learning;* Gabel, D., Ed.; MacMillan: New York, NY, USA, 1994; p. 327.
[76] Heller, P.; Hollabaugh, M. Teaching problem solving through cooperative grouping. Part 1. Group versus individual problem solving. *Am. J. Phys.* **1992**, *60*, 627.
[77] Heller, P., & Hollabaugh, M. Teaching problem solving through cooperative grouping. Part 2. Designing problems and structuring groups. *Am. J. Phys.* **1992**, *60*, 637.
[78] Zhang, J.; Norman, D. Representations in distributed cognitive tasks. *Cog. Sci.* **1994**, *18*, 87-122.
[79] Zhang, J.; Patel, V. Distributed cognition, representation and affordance. *Prag. Cog.* **2006**, *14*, 333-341.

[80] Zhang, J. The nature of external representations in problem solving. *Cog. Sci.* **1997**, *21*, 179-217.
[81] Singh, C. Impact of peer interaction on conceptual test performance. *Am. J. Phys.* **2005**, *73*, 446-451.
[82] Johnson, D.; Johnson, R. An educational psychology success story: Social interdependence theory and cooperative learning. *Educ. Res.* **2009**, *38*, 365-379.
[83] Van Heuvelen, A. Overview, case study physics. *Am. J. Phys.* **1991**, *59*, 898-907.
[84] Ward, M.; Sweller, J. Structuring effective worked examples. *Cog. Instruct.* **1990**, *7*, 1–39.
[85] Schoenfeld, A.; Herman., D. Problem perception and knowledge structure in expert and novice problem solvers. *J. Exp. Psych.: Learning, Memory and Cognition* **1982**, *8*, 484-494.
[86] Singh, C. Categorization of problems to assess and improve proficiency as teachers and learners. *Am. J. Phys.* **2009**, *77,* 73-80.
[87] Mason, A.; Singh, C. Assessing expertise in introductory physics using categorization task. *Phys. Rev. ST-PER* **2011**, *7*, 020110.
[88] Lin, S.-Y.; Singh, C. Categorization of quantum mechanics problems by professors and students. *Eur. J. Phys.* **2010**, *31*, 57-68 (2010).
[89] Docktor, J. et al. Impact of a short intervention on novices' categorization criteria. *Phys. Rev. ST-PER* **2012**, *8*, 020102.
[90] Mason, A.; Singh, C. Using categorization of problems as an instructional tool to help introductory students learn physics, *Phys. Educ.* **2016**, *51*, 025009.
[91] Singh, C. Problem solving and learning. *AIP Conf. Proc.* **2009**, *1140*, 183-197.
[92] Clement, J. Students' preconceptions in mechanics. *Am. J. Phys.* **1982**, *50,* 66-71.
[93] Posner, G. J.; Strike, K. A.; Hewson, P. W.; Gertzog, W. A. Accommodation of a scientific conception: Toward a theory of conceptual change. Sci. Educ. **1982**, *66*, 211-227.
[94] Ginsberg, H.; Opper, S. *Piaget's Theory of Intellectual Development.* Prentice Hall: Englewood Cliffs, NJ, USA, 1969.
[95] de Groot, A. Perception and memory versus thought: Some old ideas and recent findings. In *Problem solving*; Kleinmuntz, B., Ed; Wiley: New York, NY, USA, 1966; p. 19.
[96] Polya, G. How to Solve It. Princeton University Press: Princeton, NJ, USA, 1973.
[97] Schoenfeld. A. Teaching problem-solving skills. *Am. Math. Mon.* **1980**, *87*, 794-805.
[98] Reif, F.; Scott, L. Teaching scientific thinking skills: Students and computers coaching each other. *Am. J. Phys.* **1999**, *67*, 819-831.
[99] Harper, K. Student problem-solving behaviors. *Phys. Teach.* **2006**, *44*, 250-251.
[100] Maloney, D. An overview of physics education research on problem solving. In *Getting Started in PER*; American Association of Physics Teachers: College Park, MD, USA, 2011; p. 1.
[101] Huffman, D. Effect of explicit problem solving strategies on high school students' problem-solving performance and conceptual understanding of physics. *J. Res. Sci. Teach.* **1997**, *34*, 551-570.
[102] Hsu, L. et al. Resource Letter RPS-1: Research in problem solving. *Am. J. Phys.* **2004**, *72*, 1147-1156.
[103] Singh, C. Effect of misconception on transfer in problem solving. *AIP Conf. Proc.* **2007**, *951*, 196-199.
[104] Podolefsky, N.; Finkelstein, N. Analogical scaffolding and the learning of abstract ideas in physics: Empirical studies. *Phys. Rev. ST-PER* **2007**, *3*, 020104.
[105] Koedinger, K.; Nathan, M. The real story behind story problems: Effects of representations on quantitative reasoning. *J. Learn. Sci.* **2009**, *13*, 129-164.
[106] Kohl, P. et al. Strongly and weakly directed approaches to teaching multiple representation use in physics. *Phys. Rev. ST-PER* **2007**, *3*, 010108.
[107] Lin, S.-Y.; Singh, C. Challenges in using analogies. *Phys. Teach.* **2011**, *49*, 512-513.
[108] Lin, S.-Y.; Singh, C. Using isomorphic problems to learn introductory physics. *Phys. Rev. ST-PER* **2011**, *7*, 020104.
[109] Maries, A. et al. Challenges in designing appropriate scaffolding to improve students' representational consistency: The case of a Gauss's law problem. *Phys. Rev. PER* **2017**, *13*, 020103.
[110] Mateycik, F.; Jonassen, D.H.; Rebello, N.S. Using Similarity Rating Tasks to Assess Case Reuse in Problem Solving. *AIP Conf. Proc.* **2009**, 1179, 201-204.
[111] Gladding, G. et al. Clinical study of student learning using mastery style versus immediate feedback online activities. *Phys. Rev. PER* **2015**, *11*, 010114.
[112] Schroeder, N. et al. Narrated animated solution videos in a mastery setting. *Phys. Rev. PER* **2015**, *11*, 010103.
[113] Ryan, Q. et al. Computer problem-solving coaches for introductory physics: Design and usability studies. *Phys. Rev. PER* **2016**, *12*, 010105.
[114] Gutmann, B. et al. Mastery-style homework exercises in introductory physics courses: Implementation matters. *Phys. Rev. PER* **2018**, *14*, 010128.
[115] Morphew, J. et al. Effect of presentation style and problem-solving attempts on metacognition and learning from solution videos. *Phys. Rev. PER* **2020**, *16*, 010104.
[116] Larkin, J. The role of problem representation in physics. In *Mental Models*; Gentner, D., Stevens A., Eds.; Lawrence Erlbaum Associates: Hillsdale, NJ, 1983; p. 75.
[117] Meltzer, D. Relation between students' problem solving performance and representational mode. *Am. J. Phys.* **2005**, *73*, 463-478.
[118] Heckler, A. Some consequences of prompting novice physics students to construct force diagrams. *Int. J. Sci. Educ.* **2010**, *32*, 1829-1851.

[119] Bajracharya, R. et al. Students' strategies for solving a multirepresentational partial derivative problem in thermodynamics. *Phys. Rev. PER* **2019**, *15*, 020124.
[120] Vignal, M.; Wilcox, B. Investigating unprompted and prompted diagrams generated by physics majors during problem solving. *Phys. Rev. PER* **2022**, *18*, 010104.
[121] Zimmerman, B. Self-efficacy: An essential motive to learn. *Contemp. Educ. Psych.* **2000**, *25*, 82.
[122] Pintrich, P. A motivational science perspective on the role of student motivation in learning and teaching contexts. *J. Exp. Psych.* **2003**, *95,* 667-686.
[123] Bandura, A. *Self-efficacy: The Exercise of Control.* Freeman: New York, NY, USA, 1997.
[124] Mason, A.; Singh, C. Do advanced students learn from their mistakes without explicit intervention? *Am. J. Phys.* **2010**, *78*, 760-767.
[125] Yerushalmi, E. et al. What do students do when asked to diagnose their mistakes? Does it help them? I. An atypical quiz context. *Phys. Rev. ST-PER* **2012**, *8*, 020109.
[126] Yerushalmi, E. et al. What do students do when asked to diagnose their mistakes? Does it help them? II. A more typical quiz context. *Phys. Rev. ST-PER* **2012**, *8*, 020110.
[127] Brown, B. et al. Improving performance in quantum mechanics with explicit incentives to correct mistakes. *Phys. Rev. PER* **2016**, *12,* 010121.
[128] Singh, C.; Haileselassie, D. Developing problem solving skills of students taking introductory physics via web-based tutorials. *J. Coll. Sci. Teach.* **2010**, *39*, 34-41.
[129] Marshman, E. et al. Challenge of helping introductory physics students transfer their learning by engaging with a self-paced learning tutorial. *Frontiers in Science* **2018**, *5*, 1-15.
[130] Marshman, E. et al. Holistic framework to help students learn effectively from research-validated self-paced learning tools. *Phys. Rev. PER* **2020**, *16*, 020108.
[131] Koenig, K. et al. Promoting problem solving through interactive video-enhanced tutorials. *Phys. Teach.* **2022**, *60*, 331-334.
[132] Justice, P. Helping Students Learn Quantum Mechanics using Research-Validated Learning Tools. PhD Dissertation, 2019.
[133] McDaniel, M. et al. Dissociative conceptual and quantitative problem solving outcomes across interactive engagement and traditional format introductory physics. *Phys. Rev. PER* **2016**, *12*, 020141.
[134] Singh, C. Coupling conceptual and quantitative problems to develop student expertise in introductory physics. *AIP Conf. Proc.* **2008**, *1064*, 199-202.
[135] Mazur, E. *Peer Instruction: A User's Manual.* Prentice-Hall: Engelwood Cliffs, NJ, USA, 1997.
[136] McDermott, L. Oersted Medal Lecture 2001: Physics education research–The key to student learning. *Am. J. Phys.* **2001**, *69,* 1127-1137.
[137] Shulman, L. Those who understand: Knowledge growth in teaching. *Educ. Res.* **1986**, *15* 4-14.
[138] Shulman, L. Knowledge and teaching: Foundations of the new reform. *Harvard Educ. Rev.* **1987**, *57*, 1-22.
[139] Vygotsky, L. *Mind in Society: The Development of Higher Psychological Processes.* Harvard University Press: Cambridge, MA, USA, 1978.
[140] Docktor, J. et al. Assessing student written problem solutions: A problem-solving rubric with application to introductory physics. *Phys. Rev. PER* **2016**, *12*, 010130.
[141] Reif, F.; Larkin, J. Cognition in scientific and everyday domains: Comparison and learning implications. *J. Res. Sci. Teach.* **1991**, *28*, 733-760.
[142] Palincsar, A.; Brown, A. Reciprocal teaching of comprehension-monitoring activities, *Technical Report No. 269*, **1983**, Center for the Study of Reading.
[143] Mason, A. et al. Learning from mistakes: The effect of students' written self-diagnoses on subsequent problem solving. *Phys. Teach.* **2016**, *54*, 87-90.
[144] Holyoak, K. The pragmatics of analogical transfer. In *The Psychology of Learning and Motivation*; Bower, G.H., Ed.; Academic Press: New York, NY, 1985; p. 59.
[145] Novick, L. Analogical transfer, problem similarity, and expertise*. J. Exp. Psych.: Learning, Memory, and Cognition* **1988**, *14*, 510–520.
[146] Clement, J. Observed methods for generating analogies in scientific problem solving. *Cog. Sci.* **1998**, *12*, 563-586.
[147] Singh, C. Assessing student expertise in introductory physics with isomorphic problems, Part I: Performance on a non-intuitive problem pair from introductory physics. *Phys. Rev. ST-PER* **2008**, *4*, 010104.
[148] Singh, C. Assessing student expertise in introductory physics with isomorphic problems, Part II: Examining the effect of some potential factors on problem solving and transfer. *Phys. Rev. ST-PER* **2008**, *4*, 010105.
[149] Mason, A.; Singh, C. Helping students learn effective problem solving strategies by reflecting with peers. *Am. J. Phys.* **2010**, *78*, 748-754.
[150] Mason, A.; Singh, C. Impact of guided reflection with peers on the development of effective problem solving strategies and physics learning. *Phys. Teach.* **2016**, *54*, 295-299.
[151] Badeau, R. et al. What works with worked examples: Extending self-explanation and analogical comparison to synthesis problems. *Phys. Rev. PER* **2017**, *13*, 020112.
[152] Ibrahim, B. et al. How students process equations in solving quantitative synthesis problems? Role of mathematical complexity in students' mathematical performance. *Phys. Rev. PER* **2017**, *13*, 020120.
[153] Ibrahim, B. et al. Students' conceptual performance on synthesis physics problems with varying mathematical complexity. *Phys. Rev. PER* **2017**, *13*, 010133.

[154] Ibrahim, B.; Ding, L. Sequential and simultaneous synthesis problem solving: A comparison of students' gaze transitions, *Phys. Rev. PER* **2021**, *17*, 010126.

[155] Al-Salmani, F., & Thacker, B. Rubric for assessing thinking skills in free-response exam problems. *Phys. Rev. PER* **2021**, *17*, 010135.

[156] Burkholder, E. et al. Template for teaching and assessment of problem solving in introductory physics. *Phys. Rev. PER* **2020**, *16*, 010123.

[157] Maries, A., & Singh, C. Helping students become proficient problem solvers Part II: An example from waves. *Sci. Educ.* **2023**, *13*(1)*, xx.*

[158] Maries, A.. et al. Active Learning in an Inequitable Learning Environment Can Increase the Gender Performance Gap: The Negative Impact of Stereotype Threat. *Phys. Teach.* **2020**, *58*, 430.